\documentclass[12pt]{article}

\usepackage{amsmath}
\usepackage{amssymb}
\usepackage{dsfont} 

\usepackage{graphicx}

\usepackage{color}

\newcommand{\be}{\begin{equation}}
\newcommand{\ee}{\end{equation}}
\newcommand{\bel}[1]{\begin{equation}\label{#1}}
\newcommand{\bea}{\begin{eqnarray}}
\newcommand{\eea}{\end{eqnarray}}
\newcommand{\balign}{\begin{align}}
\newcommand{\ealign}{\end{align}}
\newcommand{\ba}{\begin{array}}
\newcommand{\ea}{\end{array}}
\newcommand{\bfig}{\begin{figure}}
\newcommand{\efig}{\end{figure}}

\newcommand{\eref}[1]{(\ref{#1})}

\newcommand{\Fref}[1]{Fig.~\ref{#1}}

\allowdisplaybreaks

\newcommand{\bra}[1]{\mbox{$\langle \, {#1}\, |$}}
\newcommand{\ket}[1]{\mbox{$| \, {#1}\, \rangle$}}
\newcommand{\exval}[1]{\mbox{$\langle \, {#1}\, \rangle$}}
\newcommand{\inprod}[2]{\mbox{$\langle \, {#1} \, | \, {#2} \, \rangle$}}

\newcommand{\n}{\hat{n}}

\newcommand{\rmd}{\mathrm{d}}
\newcommand{\rme}{\mathrm{e}}

\newcommand{\ddt}{\frac{\rmd}{\rmd t}}

\newcommand{\half}{\frac{1}{2}}

\newcommand{\Tr}{\mathop{\mathrm{Tr}}\nolimits}

\newcommand{\eps}{\varepsilon}

\newcommand{\comm}[2]{\mbox{$[{#1},\,{#2}]$}}

\newcommand{\R}{{\mathbb R}}

\begin{document}

\title{Solution of the Lindblad equation for spin helix states}

\author{V. Popkov$^{1}$ \and G.M.~Sch\"utz$^{2}$}

\maketitle

{\small
\noindent $^{~1}$Helmholtz-Institut f\"ur Strahlen-und Kernphysik,
Universit\"at Bonn, Nussallee 14-16, 53119 Bonn, Germany
\\
\noindent Email: popkov@uni-bonn.de

\smallskip
\noindent $^{~2}$Institute of Complex Systems II,
Forschungszentrum J\"ulich, 52425 J\"ulich, Germany
\\
\noindent Email: g.schuetz@fz-juelich.de
}

\begin{abstract}
Using Lindblad dynamics we study quantum spin systems with dissipative 
boundary dynamics that generate a stationary nonequilibrium 
state with a non-vanishing spin current that is locally conserved except at the
boundaries. We demonstrate that with suitably
chosen boundary target states one can solve the many-body Lindblad
equation exactly in any dimension. As solution we obtain pure states at any finite 
value of the dissipation
strength and any system size. They are characterized by a helical stationary 
magnetization profile and a superdiffusive ballistic current of order one,
independent of system size even when the quantum spin system is not integrable. 
These results are derived in explicit form for the
one-dimensional spin-1/2 Heisenberg chain and its higher-spin generalizations
(which include for spin-1 the integrable Zamolodchikov-Fateev model and the 
bi-quadratic Heisenberg chain). The extension of the results to higher dimensions
is straightforward. 
\end{abstract}


\section{Introduction}

A question of considerable interest in the context of one-dimensional
transport phenomena is the magnitude of stationary
currents in boundary-driven quantum spin systems
as a function of system size $N$. In the case of normal (diffusive)
transport a current $j$ is asymptotically proportional
to $1/N$, while for ballistic transport the current approaches
a non-zero constant even in the thermodynamic
limit $N \to \infty$. In one dimension this behavior is a hallmark
of integrable systems and manifests itself in a finite Drude weight
\cite{Sirk09,Pros11a}.
A way to measure this quantity experimentally in such systems has been
proposed recently \cite{Karr16}.

We address the relationship between the nature of the boundary driving, integrability
and transport properties by studying boundary-driven quantum spin chains 
in the by now theoretically well-established and experimentally
accessible framework of non-equilibrium Lindblad dynamics. This approach models
a dissipative coupling of a quantum system to its environment and thus
allows for the description of stationary current-carrying quantum states.
We explore conditions on the boundary driving under which
ballistic transport may occur in a
quantum spin system. It turns out that such behavior arises
in stationary states in which
the ballistic current is associated with a
spin rotation along the direction of driving.
We shall call such superdiffusive nonequilibrium stationary
states ``spin helix states'' (SHS), in analogy to phenomena in
spin-orbit-coupled two-dimensional electron systems
\cite{Bern06,Liu06,Kora09}. 
We focus on one-dimensional spin chains, which are of great current interest.
However, it will transpire that analogous SHS will appear also in higher
dimensions with an appropriate choice of Lindblad boundary driving.

The 1-d SHS generalizes the
asymptotic state in the isotropic Heisenberg chain ($XXX$-chain)
in the thermodynamic limit $N\to\infty$ that was found recently
\cite{Kare13a,Kare13b} which is, in turn, reminiscent of the
helical ground state of the {\it classical} isotropic Heisenberg spin chain
with boundary fields and its formal analog of
ferromagnetic quantum domains in the
Heisenberg quantum chain \cite{Schi77,Alca95}.
The novelty of the SHS is the occurrence of a non-zero winding number in the
helical state that turns out to be responsible for the ballistic transport.

Mainly we are interested
in exact SHS's in the experimentally relevant chains of finite
length. However, we shall also present numerical results away from the
exactly solvable points that highlight the specific features of the exact
SHS. Interestingly, these SHS
are pure states, which is unusual for
solutions of a many-body Lindblad equation.
These states arise in the regime
$|\Delta|<1$ for the anisotropy parameter of the spin-$s$ chain.
For the ground state of the spin-1/2 XXZ Heisenberg chain this is the
quantum critical regime, unlike the ferromagnetic regime $\Delta \geq 1$
studied in \cite{Alca95}, which exhibits a mathematically somewhat analogous
but physically very different behavior.
 Notice that the nonequilibrium stationary
state of a dissipatively
boundary driven $XXZ$-chain
 was argued to converge to
the SHS  
in the Zeno limit of infinitely large boundary dissipation \cite{PopkovPresilla2016,PopkovPresillaSchmidt2017}.
Here we show how the SHS is  produced at arbitrary \textit{finite} dissipative strength.

The paper is organized as follows. To be
concrete, we first consider in Sec. 2 the anisotropic
spin-1/2 Heisenberg chain. We define the
SHS and derive the conditions under which exact
SHS's arise with judiciously chosen Lindblad dissipators. In Sec. 3
we discuss in some detail transport properties of the
spin-1/2 SHS and compare with transport in non-SHS states.
Then we go on to generalize the approach to higher-spin chains (Sec. 4)
and discuss some classical analogies.
In Sec. 5 we draw some conclusions.

\section{Spin helix states in the spin-1/2 $XXZ$-chain}

The spin-1/2 $XXZ$-chain is defined by the Hamiltonian \cite{Baxt82}
\bel{Def:XXZ}
H = \sum_{k=1}^{N-1} h_k
\ee
with local interaction matrices $h_k$ given in terms of Pauli spin-1/2
matrices by
\bea
\label{Hlocal1}
h_k & = & J\left[\sigma^x_{k} \sigma^x_{k+1} + \sigma^y_{k} \sigma^y_{k+1} +
\Delta (\sigma^z_{k} \sigma^z_{k+1} - 1)\right] \\
\label{Hlocal2}
& = & 2J \left[ \sigma^+_{k} \sigma^-_{k+1} + \sigma^-_{k} \sigma^+_{k+1}
- \cos{\eta} \left( \n_{k} \hat{v}_{k+1} + \hat{v}_{k} \n_{k+1}
\right) \right].
\eea
Here $\Delta = \cos{\eta}$ is the anisotropy parameter,
and in the second representation we have used the local projectors
\bel{nkvk}
\n_{k} = \frac{1}{2} \left(1-\sigma^z_k\right), \quad
\hat{v}_{k} = \frac{1}{2} \left(1+\sigma^z_k\right)
\ee
and the spin raising and lowering operators
$\sigma_k^\pm = (\sigma_k^x \pm i \sigma_k^y)/2$. We recall that
the Pauli matrices satisfy the $SU(2)$ commutation relations
$\comm{\sigma^\alpha_k}{\sigma^\beta_l} = 2 i \delta_{k,l} \sum_{\gamma=1}^{3}
\epsilon_{\alpha\beta\gamma} \sigma^\gamma_k$ where
$\epsilon_{\alpha\beta\gamma}$ is the totally antisymmetric Levi-Civita
symbol with $\epsilon_{123}=1$.

The object of interest is the density matrix $\rho$ in a
boundary-driven non-equilibrium situation where stationary
currents arise from the coupling of the left and right boundary sites
$1$ and $N$ to an environment which projects the boundary spins in different
directions.
The density matrix $\rho$ of the
non-equilibrium steady state (NESS) is
determined by the stationary Lindblad equation \cite{Breu02,Atta06}
\bel{Def:Lindblad}
0 = \ddt \rho = -i\comm{H}{\rho} + \mathcal{D}_L(\rho) + \mathcal{D}_R(\rho)
\ee
with boundary dissipators $\mathcal{D}_{j}$, $j\in\{L,R\}$ acting on the
density matrix as
\bel{Def:dissipatorgeneral}
\mathcal{D}_j(\rho) = D_j \rho D_j^\dagger  - \half \{D_j^\dagger D_j, \rho \}.
\ee
The Lindblad operators $D_j$ which encode the nature of the
boundary driving will be specified below.
Stationary expectations $\exval{O}$ of physical observables $O$ are then
given by the trace $\exval{O} = \Tr(O\rho)$. Our main interest will be
in the magnetic moments $\vec{m}_k$ at site $k$ of the chain.
For convenience we ignore material-dependent factors and
choose units such that $\vec{m}_k = \exval{\vec{\sigma}_k}$.

In the absence of the unitary part given by the spin chain
Hamiltonian $H$, the non-unitary
dissipative part given by the dissipators $\mathcal{D}_{j}$
forces the system locally at the respective left (L) or right (R)
boundary site into some target state. Thus, if the two
target states are different,
stationary currents associated with local bulk-conserved degrees of freedom
are generally expected to flow due to the action of the unitary bulk part of
the Lindblad equation.

\subsection{The spin-1/2 helix state}

For many problems of interest the quantum master equation
\eref{Def:Lindblad} admits an exact solution
in which the stationary density matrix is expressed in matrix product form
\cite{Pros15,Kare16}. Here we take a different approach and make a
pure-state ansatz
\bel{densitymatrix}
\rho = \ket{\Phi}\bra{\Phi}
\ee
with the product state
\bel{Phi}
\ket{\Phi} = \ket{\phi_1}\otimes \dots \otimes \ket{\phi_N}.
\ee
This means that we can write
\bel{densitymatrix2}
\rho = \ket{\phi_1}\bra {\phi_1} \otimes \dots \otimes \ket{\phi_N}\bra {\phi_N}.
\ee
We take the basis where the $z$-components $\sigma_k^z$ of the
local spin operator are all diagonal and choose
\bel{philocal}
\ket{\phi_k} = \frac{1}{\sqrt{|a|^2+|b|^2}}
\left( \ba{c} a \, \rme^{-i\half \phi_k} \\ b \, \rme^{i\half \phi_k} \ea \right)
\ee
with the local phase angle
\bel{phik}
\phi_k = \varphi k
\ee
where $0 \leq \varphi < 2\pi$.

With the
parametrization $a = \rme^{i\varphi_B/2}$, $b = r \rme^{-i\varphi_B/2}$
the magnetization profiles $m^\alpha_k :=
\exval{\sigma^\alpha_k}/2$, i.e., the $\alpha$-components of the dimensionless
magnetic moments,
are given by
\bel{spin}
m^x_k
= \frac{r}{1+r^2} \cos{(\varphi k - \varphi_B)}, \quad
m^y_k
= \frac{r}{1+r^2} \sin{(\varphi k - \varphi_B)}, \quad
m^z_k = \half \frac{1-r^2}{1+r^2}.
\ee
One recognizes in $\varphi$ the twist angle between neighbouring spins
in the $xy$-plane.
Therefore we refer to the pure density matrix
\eref{densitymatrix2} specified by the properties \eref{philocal}
and \eref{phik} as spin helix state (SHS).

The quantity
$\varphi (N-1)$ yields the twist angle between boundary target polarizations
in the $xy$-plane. Hence any $\varphi \in [0,2\pi[$ of the form
\bel{gamma}
\varphi = \frac{\Phi + 2\pi K}{N-1}
\ee
with $0 \leq \Phi < 2\pi$ and $0 \leq K < N-1$
gives rise to the same spin rotation between the boundary spins by the
angle $\Phi$ in the $xy$-plane.
We shall refer to $\Phi$ as the boundary twist
and to $K$ as the (clockwise) winding number of the spin helix \cite{WindingNumber}. Without loss of generality
we fix the phase $\varphi_B=\varphi$ which corresponds to a choice of the
coordinate system such that the planar spin component at site 1 points into
the $x$-direction. The left target state at site 1 is then the local
density matrix $\rho_L = (\hat{v} + r^2 \hat{n} + r\sigma^x)/(1+r^2)$.
and the right target state is given by $\rho_R = (\hat{v} + r^2 \hat{n} +
r\cos{(\Phi)}\sigma^x + r\sin{(\Phi)}\sigma^y)/(1+r^2)$.
For $r=1$ the SHS is fully
polarized in the $xy$-plane with perpendicular magnetization $m^z_k=0$
along the chain.
Due to the factorized structure of the SHS there are no spin-correlations
between different sites.

Thermal-like properties of this NESS can be characterized by the
bond energy density $\varepsilon_k := \exval{h_k}$. From the
factorization property \eref{densitymatrix2} and the explicit form
of the local magnetizations \eref{spin} one finds that the
bond energy density is spatially constant and given by
\bel{energy}
\varepsilon  = J \left[ \left(\frac{2r}{1+r^2}\right)^2 \cos{\varphi} + \Delta
\left( \left(\frac{1-r^2}{1+r^2}\right)^2 - 1 \right) \right] .
\ee
Due do the factorized structure of the SHS there are no energy correlations
between non-neighbouring bonds.

The complete absence of correlations in the SHS is reminiscent of very
high temperatures.
We caution, however, not to interpret this lack of correlations and
the flat energy profile along the chain as indicating
proximity to some equilibrium state $\rho \propto \exp{(-\beta_{eff} H)}$
with an effective temperature given by \eref{energy}, not even if $\varphi=0$
when also the magnetization profile is flat.
For $\varphi=0$ one can write $\rho \propto \exp{(-\beta_{eff} H_{eff})}$
with an effective Hamiltonian of the form
$H_{eff} = \sum_k (\sigma_k^z + u \sigma_k^x + w)$.
Such a non-interacting Hamiltonian corresponds to a subspace of $H$
for $\Delta=0$ \cite{Schu95}, but does not in general capture any significant
physical property of the thermal density matrix
$\rho \propto \exp{(-\beta H)}$ for any finite temperature at any value of
$\Delta$.

\subsection{Construction of the boundary dissipators}

Now we aim at deriving boundary dissipators which allow for
maintaining the SHS stationary in the {\it finite} $XXZ$-chain. To this end
we first make a remark on pure-state solutions of
a general stationary Lindblad equation
\bel{Def:Lindbladgeneral}
\mathcal{L}(\rho) = -i\comm{H}{\rho} + \sum_j \mathcal{D}_j(\rho) = 0
\ee
where here $j$ belongs to some index set (not necessarily just $L$ and $R$).
Let a pure state
$\rho = \ket{\Psi}\bra{\Psi}$ be the solution of \eref{Def:Lindbladgeneral}.
Then $\ket{\Psi}$ is an eigenvector of all the Lindblad operators $D_j$ and the
Lindblad equation turns into the set of eigenvalue problems
\bel{Lindbladpure}
D_j \ket{\Psi} = \lambda_j \ket{\Psi}, \quad \tilde{H} \ket{\Psi} = \mu \ket{\Psi}
\ee
with (in general complex) eigenvalues $\lambda_j$ and (real) eigenvalue $\mu$ of
the shifted Hamiltonian
\bel{Hshift}
\tilde{H} = H + \sum_j \frac{i}{2}
\left( \bar{\lambda}_j D_j - \lambda_j D^\dagger_j\right).
\ee

This can be seen as follows \cite{Yamamoto05,ZollerPRA08}.
Sandwich the Lindblad equation \eref{Def:Lindbladgeneral} with
$\ket{\Psi}$. Then the unitary part involving the commutator with $H$
vanishes identically and one gets
\be
\sum_j \left(\bra{\Psi} D_j \ket{\Psi} \bra{\Psi} D_j^\dagger \ket{\Psi} -
\bra{\Psi} D_j^\dagger D_j \ket{\Psi}\right) = 0
\ee
for the dissipative part.
By the Schwarz inequality (which generally gives $\geq 0$ for the l.h.s.)
the equality is realized if and only if the eigenvalue property
\bel{Lindbladeigen}
D_j \ket{\Psi} = \lambda_j \ket{\Psi}
\ee
holds for each dissipative term.
Then the Lindblad dissipator can be written as a commutator
\bel{disscomm}
\mathcal{D}_j(\rho) =
\half \lambda_j \comm{\rho}{D_j^\dagger} + \half \bar{\lambda}_j \comm{D_j}{\rho}
= \comm{\half ( \bar{\lambda}_j D_j - \lambda_j D_j^\dagger )}{\rho}
\ee
and the Lindblad equation becomes
\be
\comm{H + \sum_j \frac{i}{2} ( \bar{\lambda}_j D_j - \lambda_j D_j^\dagger )}{\rho}=0.
\ee

Consider now the commutator $\comm{A}{\sigma}=0$ with a general tensor matrix
$\sigma= \ket{\Psi}\bra{\Psi'}$ such that $\inprod{k}{\Psi}\neq 0$ and
$\inprod{\Psi'}{l}\neq 0$ for all orthonormal basis vectors $\ket{k}$,
$\ket{l}$
of the separable
Hilbert space to which $\ket{\Psi}$ and $\ket{\Psi'}$ belong.
Sandwiching with $\bra{k}$ and $\ket{l}$ yields
\be
\bra{k} A \ket{\Psi} \inprod{\Psi'}{l} = \inprod{k}{\Psi} \bra{\Psi'} A \ket{l}
\ee
or, equivalently,
\be
\frac{\bra{k} A \ket{\Psi}}{\inprod{k}{\Psi}} =
\frac{\bra{\Psi'} A \ket{l}}{\inprod{\Psi'}{l}}
\quad \forall k,l.
\ee
Hence
\be
\bra{k} A \ket{\Psi} = \mu \inprod{k}{\Psi}, \quad
\bra{\Psi'} A \ket{k} = \mu \inprod{\Psi'}{k} \quad \forall k
\ee
with the same constant $\mu$. This implies
\be
A \ket{\Psi} = \mu \ket{\Psi}, \quad \bra{\Psi'} A = \mu  \bra{\Psi'}.
\ee
This proves \eref{Lindbladpure} for any pure state.
Conversely, if \eref{Lindbladpure} holds for some vector $\ket{\Psi}$
then the pure state $\rho = \ket{\Psi}\bra{\Psi}$ is a solution of the
original Lindblad equation \eref{Def:Lindbladgeneral}.

Now we apply this property to the SHS defined by
\eref{densitymatrix2} with \eref{philocal}, \eref{phik}
which we require to satisfy the stationarity condition \eref{Def:Lindblad}
with boundary Lindblad operators $D_{L,R}$.
Notice that one can write the interaction terms $h_k$ of
the $XXZ$-Hamiltonian \eref{Def:XXZ} as
\be
\label{Hlocal3}
h_k = e_k(\eta) + i \sin{\eta} (\sigma^z_{k+1} - \sigma^z_{k})
= e_k(-\eta) - i \sin{\eta} (\sigma^z_{k+1} - \sigma^z_{k})
\ee
with
\be
e_k(\eta) =
2 J \left( \sigma^+_{k} \sigma^-_{k+1} + \sigma^-_{k} \sigma^+_{k+1}
- \rme^{i\eta} \n_{k} \hat{v}_{k+1} - \rme^{-i\eta} \hat{v}_{k} \n_{k+1} \right).
\ee
This fact allows us to write
\bel{H2}
H = G(\eta) + i J \sin{\eta} (\sigma^z_{N} - \sigma^z_{1})
=  G(-\eta) - i J \sin{\eta} (\sigma^z_{N} - \sigma^z_{1})
\ee
with
$G(\eta) = \sum_{k=1}^{N-1} e_k(\eta)$.

Remarkably, for the relation
\bel{etagamma}
\eta = \varphi
\ee
between the twist angle $\varphi$ of the SHS and the anisotropy $\eta$ of the
$XXZ$-chain
one has
\be
e_k(\varphi) \ket{\Phi} = 0, \quad \bra{\Phi} e_k(-\varphi) = 0.
\ee
This implies $G(\varphi) \ket{\Phi}=0$ and $\bra{\Phi}G(-\varphi)=0$ and therefore
\bel{telescope}
H \ket{\Phi} = i J \sin{\varphi} (\sigma^z_{N} - \sigma^z_{1}) \ket{\Phi}, \quad
\bra{\Phi} H = - i J \sin{\varphi} \bra{\Phi} (\sigma^z_{N} - \sigma^z_{1}).
\ee

To proceed and construct suitable Lindblad operators
$D_{L,R}$ it is convenient to define for subscript $j\in \{L,R\}$
the shifted Lindblad operators
\be
\label{DefTildeL}
\tilde{D}_j = D_j - \lambda_j  .
\ee
We also note that we can write the shifted Hamiltonian \eref{Hshift} as
\be
\label{DefHtilde}
\tilde{H} = H + \sum_{j\in \{L,R\}} \frac{i}{2}
( \bar{\lambda}_j \tilde{D}_j - \lambda_j \tilde{D}_j^{\dagger}).
\ee
The constants $\lambda_j$ are to be determined.
According to \eref{Lindbladpure} this implies that one has to solve
\bel{Lindbladnew1}
\tilde{D}_L \ket{\Phi} = \tilde{D}_R \ket{\Phi}=0,
\ee
and
\bel{Lindbladnew2}
\bra{\Phi} \left[ -i J \sin{(\varphi)} (\sigma^z_{N} - \sigma^z_{1}) +
\frac{i}{2}
\left( \bar{\lambda}_L \tilde{D}_L+ \bar{\lambda}_R \tilde{D}_R\right)
\right]  = \mu \bra{\Phi}
\ee
with $\mu \in \R$. Here we used that \eref{Lindbladnew1} is equivalent to
$\bra{\Phi} \tilde{D}^\dagger_i=0$.
This allows us to split these four equations into two pairs of equations
for each boundary
\bel{Lindbladleft}
\tilde{D}_L \ket{\Phi} =0, \quad
\bra{\Phi} \left(  i J \sin{(\varphi)}  \sigma^z_{1} +
\frac{i}{2} \bar{\lambda}_L \tilde{D}_L
\right)  = \mu_L \bra{\Phi}
\ee
\bel{Lindbladright}
\tilde{D}_R \ket{\Phi} = 0, \quad
\bra{\Phi} \left( -i J \sin{(\varphi)}  \sigma^z_{N} +
\frac{i}{2} \bar{\lambda}_R \tilde{D}_R
\right)  = \mu_R \bra{\Phi}
\ee
with $\mu_L = (\mu + i \nu)/2$ arbitrary and $\mu_R = \bar{\mu}_L$
so that $\mu_L + \mu_R = \mu \in \R$ as required by \eref{Lindbladpure}.
The real-valued constants $\mu, \nu$ can be computed by multiplying
from the right by $\ket{\Phi}$. Using \eref{spin} yields
\bel{mu1N}
\mu_L =  i J \sin{(\varphi)} \ \frac{1-r^2}{1+r^2} = - \mu_R
\ee
and therefore $\mu = 0$, $\nu = j^z$. For full planar polarization
this reduces to $\mu_L=\mu_R=0$.

Requiring
the left dissipator $D_L$ to act non-trivially on the left boundary site 1
one finds from
the first eigenvalue equation in \eref{Lindbladleft} that
\be
\label{L1gen}
\tilde{D}_L
= \left( \ba{cc}
r \alpha_L & -\alpha_L \\[2mm]
r \beta_L & -\beta_L \ea \right)_1
= \alpha_L \left(r \hat{v}_1 - \sigma_1^+ \right)
- \beta_L \left(\n_1 - r \sigma_1^- \right)
\ee
 with arbitrary constants
$\alpha_L, \beta_L$.
Then the second equation in \eref{Lindbladleft} is solved by
\be
 \bar{\lambda}_L = -  \frac{4 r J \sin{\varphi}}{(1+r^2)(\alpha_L +r \beta_L)} .
\label{CondEivModifiedHam}
\ee

For the right boundary the eigenvalue equation
$\tilde{D}_R \ket{\Phi} =0$ in \eref{Lindbladright} gives
\bea
\label{LNgen}
\tilde{D}_R & = & \rme^{-i \frac{(N-1)\varphi}{2} \sigma_N^z} \left( \ba{cc}
r \alpha_R &
-\alpha_R \\[2mm]
r \beta_R &
-\beta_R
\ea \right)_N \rme^{i \frac{(N-1)\varphi}{2} \sigma_N^z} \nonumber \\
& = & \alpha_R \left(r \hat{v}_N - \rme^{-i \Phi} \sigma_N^+ \right)
- \beta_R \left(\n_N - r \rme^{i \Phi} \sigma_N^- \right)
\eea
with arbitrary constants $\alpha_R, \beta_R$. From the
second equation in \eref{Lindbladright} one then obtains
 \be
\bar{\lambda}_R = \frac{4 r J \sin{\varphi}}{(1+r^2)(\alpha_R +r \beta_R)} .
\label{CondEivModifiedHamR}
\ee
Thus the SHS is stationary under the action of a two-parameter
family of boundary dissipators with Lindblad operators $D_j = \tilde{D}_j + \lambda_j$.

\section{Transport properties of the SHS}

We treat both spin and energy transport, the emphasis being on spin transport.

\subsection{Spin transport in the SHS}

The $z$-component of the total magnetization is conserved under the
unitary part of the time evolution. The associated conserved spin current is
defined by the continuity equation
through the time derivative of the magnetization profile
$\dot{m}^z_k = j^z_{k-1} - j^z_k$. Since
$\dot{m}^z_k = i \exval{\comm{H}{\sigma^z_k}}/2$
one gets from the commutation relations of the Pauli matrices
the current operator
\bel{jz}
\hat{\jmath}^z_k  = J
\left(\sigma^x_{k} \sigma^y_{k+1} - \sigma^y_{k} \sigma^x_{k+1}\right).
\ee
In the stationary state
the current  $j^z := \exval{\hat{\jmath}^z_k}$ does not depend on
$k$ and it is of interest to investigate its properties in the SHS.
Strictly speaking, the SHS as defined above arises as stationary
solution of the Lindblad equation for a finite chain only in the regime
$|\Delta| < 1$
of the $XXZ$-chain. However, as shown below, it appears asymptotically
also in the isotropic Heisenberg chain with $\Delta=1$ and it has a
(non-helical) analog in the ferromagnetic regime $\Delta >1$. We discuss
these cases separately.

\subsubsection{Helical regime $|\Delta| < 1$}

The factorized form of the
SHS  defined by \eref{densitymatrix2} - \eref{phik} yields
\bel{zcurrent}
j^z = J \frac{4r^2}{(1+r^2)^2} \sin{\varphi}
\ee
which even in a large system is of order 1 for macroscopic winding numbers of order $N$.
Interestingly, in contrast to the classical relation between a locally
conserved current and boundary gradients of the associated conserved quantity,
for any winding number there is a current even though there is no gradient
$\Delta m^z := m^z_1-m^z_N =0$ between the $z$-magnetizations of the boundaries.
Moreover, the behaviour of the SHS is also
in contrast to the situation where the $XXZ$-chain is driven by {\it two}
Lindblad operators at each boundary into a state close to an
infinite-temperature
thermal state \cite{Znid11a}. In this case, the effective diffusion
coefficient $D^z_{eff} \propto  Lj^z/\Delta m^z$
was found numerically
for chains up to more than 200 sites to be proportional to $L$
(corresponding to ballistic transport) with a coefficient of proportionality
that depends on the anisotropy $\Delta$.
Theoretically, a ballistic spin current in this regime
was proved by calculating the lower bound for a respective Drude weight, see
\cite{Pros11a}.

The spin transport of the SHS is, in fact, reminiscent of the
persistent current $j$ in a mesoscopic ring threaded by a magnetic flux
$\Phi$ \cite{Byer61,Kohn64}.
At zero temperature one has
\bel{perscurr}
j = - \frac{\partial E_0 }{\partial \Phi}
\ee
and the Drude weight is given by the spin stiffness \cite{Shas90}
\bel{Drudeweight}
D = L \frac{\partial^2 E_0 }{\partial \Phi^2}|_{\Phi=\Phi_m}
\ee
where $E_0$ is the ground state energy and $\Phi_m$ is the value of
$\Phi$ that minimizes $E_0(\Phi)$. Substituting
the ground state energy $E_0$ of the ring by the energy density
\eref{energy} times the chain length $L=N-1$ (in lattice units) of the SHS,
i.e., $E_0 \to (N-1) \varepsilon$, identifying the flux $\Phi$
with the magnitude of the boundary twist, and keeping $\Delta$ fixed
when taking the derivative w.r.t. $\Phi$
one finds from \eref{perscurr} that $j = j^z$
as given by \eref{zcurrent} and then
\eref{Drudeweight} gives $D_{SHS} = |J| > 0$,
indicating infinite DC conductivity.

Expressions for {\it finite} temperature analogous to \eref{perscurr}
and \eref{Drudeweight} are derived in \cite{Cast95} and it was conjectured that a finite
Drude weight at non-zero temperature is a generic property of integrable
systems.
Thus the non-thermal (but certainly not zero-temperature)
SHS of the integrable XXZ-chain appears to fit into the picture relating
the Drude weight obtained via \eref{Drudeweight}, infinite DC conductivity
and integrability \cite{Sirk09,Pros11a,ProsIlievski13,
Pere14}. The Drude weight $D_{SHS}$, however, does not depend on the anisotropy
$\Delta$ unlike the thermal Drude weight \cite{Shas90,Benz05,Herb11}.
More significantly, however it will be shown below that the ballistic
transport in the SHS is, in fact, unrelated to integrability.

\subsubsection{Isotropic point $|\Delta| = 1$}

At the isotropic point $\Delta=1$
where $\eta=0$ and the matching condition \eref{etagamma} yields a
trivial constant SHS with twist angle $\Phi=0$ and winding number
$K=0$. However, it is interesting to look at
the magnetization profiles \eref{spin}
and the spin current \eref{zcurrent}
with the boundary driven isotropic
$XXX$-chain, corresponding to
non-zero boundary twist $\theta \neq 0$ in the
$xy$-plane. It was shown in \cite{Kare13a,Kare13b}
that the boundary target states and the magnetization
profiles for large $N$ are of the form \eref{spin}
with $\varphi = \theta/(N-1)$ and $r=1$.
Thus this non-equilibrium steady
state of the $XXX$-chain is a SHS in the thermodynamic limit with winding
number $K=0$ and boundary twist $\Phi=\theta$.

The $z$-component of the spin current in the $XXX$-chain is asymptotically
given by $j^z \sim J \theta/N$ \cite{Kare13b}, which agrees with
\eref{zcurrent} for $\varphi = \theta/(N-1)$ and large $N$ \cite{spincurrent}.
Moreover, one can show that in the $XXX$-case one has
$\Delta m^z := m^z_1 - m^z_N = O(1/N)$,
indicating ballistic
transport of the $z$-component of the spin in the $XXX$-chain since
the effective diffusion coefficient $D^z_{eff} = N j^z/(\Delta m^z)$
is proportional to system size $N$.
This is consistent with the observation of infinite conductivity in the
SHS of the XXZ-chain obtained above
from the Drude weight \eref{Drudeweight} which is finite also
for $\Delta=1$ \cite{Karr12}.

However, the ballistic transport in the SHS of the $XXX$ chain
is in contrast to the transport properties both
of the canonical ensemble for which it has been shown that
the spin stiffness of the periodic $XXX$-chain at zero $z$-magnetization
vanishes at any positive temperature \cite{Carm15} and
of the ``infinite-temperature'' $XXX$-chain with two
Lindblad operators at each boundary, reported in \cite{Znid11b}.
According to exact numerical calculations for short chains up to approx.
10 sites the diffusion coefficient seems to diverge superdiffusively
with system size as $D^z_{eff} = \propto N^{1/2}$ in this rather different
setting. This is remarkable as it implies that the microscopic details
of the Lindblad boundary dissipators may determine fundamentally qualitative
properties of the bulk.

\subsubsection{Ferromagnetic coupling $\Delta > 1$}

The Heisenberg Hamiltonian with $J<0$ and $\Delta > 1$ (corresponding
to a purely imaginary anisotropy parameter $i\eta$) has a degenerate
ferromagnetic ground state with all spins aligned in positive or negative
$z$-direction,
corresponding to the SHS with $r=0$ or $r=\infty$ respectively. We note,
however, that the
SHS with $r$ finite can be defined also for purely imaginary $\varphi$
and therefore the matching condition \eref{etagamma} can be met for $\Delta > 1$.
However, this state is not a helix state. Substituting $\varphi \to i\eta$
and parametrizing
$r=\exp{(u^\ast N \eta + i \phi_0)}$
one obtains for the Heisenberg chain \eref{Def:XXZ} with $\Delta = \cosh{\eta}$
a fully polarized state with
vanishing spin current $j^z$ and the magnetization profiles given by
\bel{SHSforImaginaryEta}
\exval{\sigma^x_k} = \frac{\cos{\phi_0}}{\cosh{(\eta\tilde{k})}}, \quad
\exval{\sigma^y_k} = \frac{\sin{\phi_0}}{\cosh{(\eta\tilde{k})}}, \quad
\exval{\sigma^z_k} = \tanh{(\eta\tilde{k})}
\ee
where $\tilde{k}=k-u_0N$.

This is the domain wall state of the XXZ-chain with opposite
boundary fields in $z$-direction
\cite{Alca95} with a left domain of negatively aligned spins and
a right domain with positively aligned spins.
For $N\gg 1/\eta^2$ the domain wall between positive and negative aligned
spins is located at $u_0N$, provided that $0 < u_0 < N$. Otherwise one has a
boundary layer with a width of order $1/\eta$.
Only in a region of size
$O(1/\eta^2)$
near the domain wall one has for large $N$ a non-negligible transverse
magnetization $m_k^{x,y}$. This domain wall state has a direct classical analog as stationary
traffic jam state of the asymmetric simple exclusion process with
reflecting boundary conditions \cite{Sand94,Schu97} since for $\Delta > 1$ the
XXZ-Hamiltonian coincides with the generator of this stochastic
interacting particle system \cite{Schu01}. Note  that also the state (\ref{SHSforImaginaryEta}) can be
dissipatively obtained for infinite dissipation strength in a $XXZ$ chain with fine-tuned
anisotropy $\Delta = \cosh \eta$ \cite{PopkovPresilla2016}.

\subsection{Energy transport in the SHS}

The operator for the locally conserved energy current $\hat{\jmath}^E_k$
associated with bond $(k,k+1)$
is defined by the continuity equation
$\dot{h}_k = i \comm{H}{h_k} = \hat{\jmath}^E_{k} - \hat{\jmath}^E_{k+1}$
which yields $\hat{\jmath}^E_{k} = i \comm{h_{k-1}}{h_k}$ \cite{Lusc76,Grab94}.
Using the
commutation relations of the Pauli matrices
one finds
\bea
\hat{\jmath}^E_k & = & 2 J^2 \left(
- \sigma_{k-1}^x\sigma_k^z\sigma_{k+1}^y
+ \Delta \sigma_{k-1}^x\sigma_k^y\sigma_{k+1}^z
+ \sigma_{k-1}^y\sigma_k^z\sigma_{k+1}^x \right. \nonumber \\
& &
\label{jE}
\left.
- \Delta \sigma_{k-1}^y\sigma_k^x\sigma_{k+1}^z
- \Delta \sigma_{k-1}^z\sigma_k^y\sigma_{k+1}^x
+ \Delta \sigma_{k-1}^z\sigma_k^x\sigma_{k+1}^y \right).
\eea
The energy current $j^E = \exval{\hat{\jmath}^E_k}$ then follows
from the factorized structure \eref{Phi} of the SHS and the magnetization
profiles \eref{spin}.

Somewhat surprisingly
\be
j^E = J^2 \frac{8r^2(1-r^2)}{(1+r^2)^3}
\left(2 \Delta \sin{\varphi} - \sin{2\varphi}\right) = 0
\ee
since $\Delta = \cos{\varphi}$ in the SHS. This is consistent with the constant bond energy along the chain (implying the absence of a energy gradient
between the boundaries), but nevertheless not completely obvious since
(a) from a microscopic perspective it is not {\it a priori} clear that the
dissipators would not generate an energy current and (b)
the total energy current $\sum_{k} \hat{\jmath}^E_{k}$ in a periodic chain
is a conserved charge of the integrable periodic $XXZ$-chain
\cite{Lusc76,Grab94} and hence ballistic transport of energy is generic.

\subsection{Numerical results}


Now we explore numerically on a concrete example
the predicted special properties of the spin helix state as opposed to a generic
non-equilibrium state that arises as a solution of the Lindblad equation
\eref{Def:Lindblad} with Lindblad operators whose parameters do {\it not}
satisfy the matching condition \eref{etagamma} and
conditions \eref{L1gen} - \eref{CondEivModifiedHamR} for the Lindblad operators.
We focus on the fully polarized SHS with $r=1$ and fix the Heisenberg exchange
coupling $J=1$.

For the numerically exact solution of the Lindblad equation
we consider an $XXZ$-chain of four sites.
For the Lindblad
operators we take $\alpha_L=\beta_L=\alpha_R=\beta_R=\sqrt{\Gamma} >0$
so that
\be
D_L  =\sqrt{\Gamma} \left( \epsilon_L  I -   \sigma^z_1 + i \sigma^y_1 \right), \quad
D_R  = \sqrt{\Gamma} \left( \epsilon_R I -  \sigma^z_N + i \cos{\Phi} \sigma^y_N
- i \sin{\Phi} \sigma^x_N \right).
\ee
For $N=4$ we take $\varphi =  2 \pi/3$ corresponding to
winding number $K=2$ and a zero
boundary twist angle $\Phi = 0$ in the $xy$-plane. By fixing
$\epsilon_R = - \epsilon_L = 0.05$ the variable $\Gamma$ becomes a measure
for the dissipative strength.
The pure SHS \eref{densitymatrix2} - \eref{phik} is then a stationary solution
of the Lindblad equation \eref{Def:Lindblad} for
\bel{SHScond}
\eta = \varphi, \quad \Gamma = \frac{\sin{\varphi}}{|\epsilon_R|}= 20 \sin{\varphi}.
\ee
For the purpose of the numerical investigation we do
{\it not} require these equations to be satisfied and
study the purity of the solution of \eref{Def:Lindblad}
and the corresponding stationary current $j^z$
as a function of the anisotropy
$\Delta = \cos{\eta}$ and the dissipative strength $\Gamma$.

As a measure for the purity of the nonequilibrium steady
state (NESS) $\rho$, we choose the von Neumann entropy
$S = -\Tr (\rho \log_2 \rho)$. Notice that $S = 0$ if and only
if the NESS is a pure state. From the exact numerical solution of
\eref{Def:Lindblad} with $\eta = \varphi$ one sees that indeed for the value of
$\Gamma$ predicted by (\ref{SHScond}) the NESS becomes pure (Fig.~1).
The spin current is maximal in amplitude near this point, but
remains approximately equally strong for all $\Gamma \gtrsim 4$.

\begin{figure}[tbp]
\centerline{\includegraphics[width=8cm,height=6cm,clip]{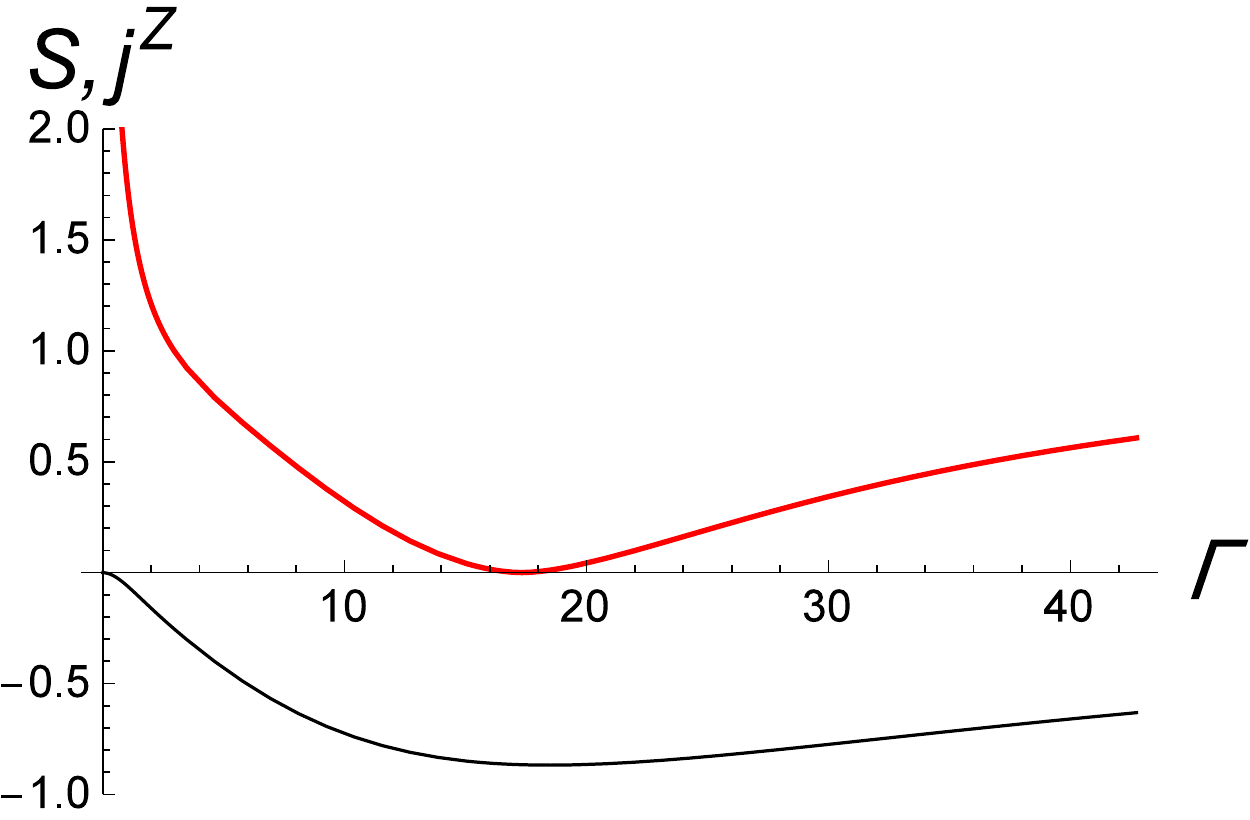}}
\caption{von-Neumann entropy $S$ (upper curve) and steady state current
$j^z$ (lower curve) versus dissipative amplitude $\Gamma$ in the
$XXZ$-chain. Parameters: $J=1, N=4, \eta=\varphi= 4 \pi/3$,
$\epsilon_R=-\epsilon_L=1/20$.
The pure state with $S=0$ describing a spin helix state is seen
for the predicted value $\Gamma= 20 |\sin{\varphi}|
\approx 17.32 $.
} \label{fig1}
\end{figure}

\begin{figure}[tbp]
\centerline{\includegraphics[width=8cm,height=6cm,clip]{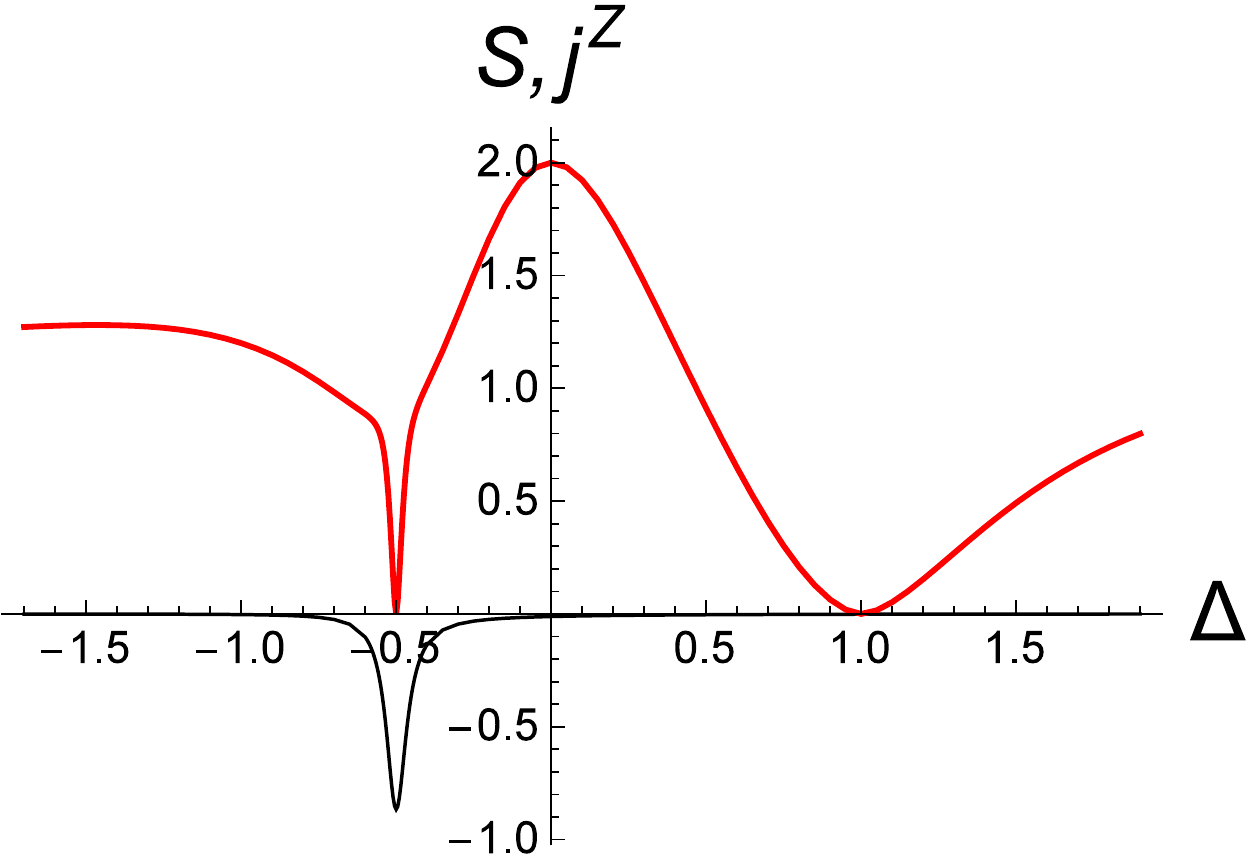}}
\caption{von-Neumann entropy $S$ (upper curve) and steady state current
$j^z$ (lower curve) versus anisotropy $\Delta$. Parameters:
$J=1, N=4, \varphi= 4 \pi/3, \Gamma =20 \sin{\varphi} $, and
$\epsilon_R= - \epsilon_L= 1/20$.
A pure SHS with $S=0$ is obtained  for the predicted value
$\Delta= \cos{\varphi} = -0.5$.
} \label{FigDeltaVarie}
\end{figure}

It is also instructive to look at the NESS as a function of the anisotropy
$\Delta = \cos{\theta}$, i.e., now we assume the dissipative strength
to satisfy (\ref{SHScond}), but not $\eta$.
In this way, we see a resonance-like behaviour
of various system observables around the critical value of the anisotropy
$\Delta=\cos{\varphi}$.  Even for a small chain of only 4 sites
the spin current $j^z$
increases by an order of magnitude and changes its sign
near the critical anisotropy, see Fig.~\ref{FigDeltaVarie}.
The von-Neumann entropy vanishes at $\Delta=\cos{\varphi}$,
as expected.
At the $XXX$-point $\Delta =1$
the von-Neumann entropy is small, but non-zero, in agreement with the
notion that the SHS is attained only asymptotically. Also the current
at this point as expected from the exact result \cite{Kare13b}.
For non-zero boundary twist $\Phi$ one obtains qualitatively similar
behavior (data not shown).

%

In order to get some insight in the resonance-like behaviour we note the following. For large amplitude  $\Gamma$, the dissipative
part of the dynamics, which is quadratic in amplitudes, becomes much larger
than the unitary Hamiltonian part  of the dynamics, and as a result the
boundary spins
$1,N$ ``freeze'' for any $\Delta$. By this we mean that the states to which
the dissipation projects the boundary spins, which are mixed states,
become very close to completely polarized pure states. At the left boundary, the
spin $1$ fixates approximately along the vector $(1,0,0)$ and at the right
boundary approximately in the direction
$(\cos \varphi (N-1),\sin \varphi (N-1),0) = (\cos{\Phi},\sin{\Phi},0)$.
Indeed, analyzing the kernel of the left  dissipator, we find that
the distance from the actually targeted state and the pure fully polarized state
at the left boundary, characterized via $\epsilon := 1-Tr(\rho_1)^2$ with
the reduced density matrix $\rho_1= Tr_{2,3,\ldots N} \rho$ is proportional
to $\eps \sim \Gamma^{-4}$ for large $\Gamma$. The same is true for the right boundary.
Now, if the polarization of the leftmost and rightmost spins in the chain
differ only slightly (in our example this boundary twist angle is actually zero
$\Phi =0$), then one expects almost no current in the system
for any $\Delta$ since
it will generically favor a homogeneous spin configuration, the
neighbouring spins at sites $k,k+1$ being almost collinear. This picture is well
borne out by \Fref{FigDeltaVarie}, except close to the critical
value $\Delta=\cos{\varphi}$. At this point the spins arrange in the
helix structure with a non-zero winding number (2 in our case)
which gives rise to the resonance.
For the exact helix spin state the spin current takes the value
$j_z= \sin \varphi \approx - 0.866$, close to the maximal possible spin
current   $|j^z_{max}|=1$.

\section{Higher-spin chains}

The above results can be  generalized to the case of spin $s$ with maximal
$z$-component $s^z=s=(n-1)/2$. We focus on spin chains with conserved $z$-component
of the total spin.

\subsection{Spin-$s$ chains with conserved $S^z$ component}

In order to define the Hamiltonian $H$ we introduce
the $n$-dimensional matrices
$E^{pq}$ with matrix elements $(E^{pq})_{mn} = \delta_{p,m}
\delta_{q,n}$. They satisfy the quadratic algbra
\bel{Ealgebra}
E^{pq} E^{p'q'} = \delta_{p'q} E^{pq'}.
\ee

From these we build the local operator
\bel{hatsz}
S^z_k := \sum_{p=0}^{2s} (s - p) E^{pp}_k
\ee
for the $z$-component of the local spin as well as the total
$z$-component
\bel{sz}
S^z := \sum_{k=1}^{N} S^z_k.
\ee

We assume a local nearest neighbour interactions
between spins, i.e.,
\begin{align}
\label{Hspin-s}
H&=\sum_{k=1}^{N-1} h_k\\
h_k &= \sum_{p,q,p',q'=0}^{2 s}
c^{pq}_{p'q'} E^{pp'}_k E^{qq'}_{k+1}
\end{align}
This notation means that the nearest neighbour interaction
matrix
\bel{localhgens}
h := \sum_{p,q,p',q'=0}^{2 s} c^{pq}_{p'q'} E^{pp'} \otimes E^{qq'}
\ee
of dimension $n^2$
has matrix elements $h_{pn+q+1,p'n+q'+1} = c^{pq}_{p'q'}$.
The coupling constants satisfy $c^{pq}_{p'q'} = \bar{c}^{p'q'}_{pq}$
since $H$ is hermitian. Moreover, we impose
the ice rule \cite{Baxt82}
\begin{align}
 c^{pq}_{p'q'} =0,\mbox { if } \ \   p+q \neq p'+ q',
 \label{icerule}
\end{align}
and the symmetry relation
\begin{align}
 c^{pq}_{p'q'} = c^{qp}_{q'p'},
 \label{refsymmetry}
\end{align}
The ice rule (\ref{icerule}) ensures conservation $\comm{H}{\hat{S}^z}=0$
of the
$z$-component of the total magnetization and (\ref{refsymmetry})
corresponds to lattice reflection symmetry $k \leftrightarrow N+1-k$.
We shall also investigate the special case of spin-flip symmetry
\bel{sfsymmetry}
c^{pq}_{p'q'} = c^{2s-p2s-q}_{2s-p'2s-q'}
\ee
which is the invariance under $S^z \leftrightarrow - S^z$. Requiring
in addition time-reversal symmetry gives the constraints
\bel{trsymmetry}
c^{pq}_{p'q'} = \bar{c}^{p'q'}_{pq}
\ee
on the phases of the coupling coefficients.

\subsection{Spin-$s$ helix state}

We
target a NESS in the form of a pure SHS $\ket{\Psi}\bra{\Psi}$
with $\ket{\Psi} = \ket{\Psi_1} \otimes \dots \otimes \ket{\Psi_N}$ and
\be
\ket{\Psi_k} = \frac{1}{\sqrt{\sum_{i=0}^{2s} |r_i|^2 }} \left(
\ba{c}
r_{0} \rme^{-i \varphi k s} \\
r_{1} \rme^{-i \varphi k (s-1)} \\
 \ldots\\
r_{2s} \rme^{i k \varphi s}
\ea
\right)
\ee
with non-zero constants $r_i$ that can be complex.
In order to achieve this state in a similar fashion
as discussed above for $s=1/2$, it is sufficient to require the generalization
 \be
 H \ket{\Psi}= (F_N-F_1)\ket{\Psi},
 \label{CondGlobalDivergence}
 \ee
of the telescopic property \eref{telescope}
 with diagonal matrices $F_k= \sum_{p=0}^{2s} f_p E^{pp}_k$.

This condition will be satisfied if
 \be
  h_k \ket{\Psi} = (F_{k+1} - F_{k}) \ket{\Psi}
  \label{CondLocalDivergence}
 \ee
 is satisfied for all $k$.
 In order to see what this implies for the
 coupling constants $c^{pq}_{p'q'}$ we define the gauge transformation
\bel{Vtrafo}
V_\varphi = \prod_{k=1}^N \rme^{i \varphi k S_k^z}
\ee
and rewrite the SHS in the form
 \be
 \ket{\Psi} = V_\varphi^{-1} \ket{\Psi_0}
\ee
where $\ket{\Psi_0}$ represents the constant wave function.
Consequently, multiplying  (\ref{CondLocalDivergence}) by
$V_\varphi$ from the left
and noting that $V_\varphi$ and $F$ are  diagonal matrices, we obtain
 \be
 V_\varphi h_k V_\varphi^{-1} \ket{\Psi_0} = ( F_{k+1} - F_{k}) \ket{\Psi_0}
  \label{CondLocalDivergenceSandwich}
 \ee
 for all $k$.
From the definition one finds $V_\varphi E_k^{pp'} V_\varphi^{-1}
= \rme^{i k (p'-p)}$ and therefore, using the ice rule,
\be
 V_\varphi h_k V_\varphi^{-1} = \sum_{p,q,p',q'=0}^{2 s}
 c^{pq}_{p'q'}
 \rme^{i \varphi (q'-q)} E^{pp'}_k E^{qq'}_{k+1}
\ee
Moreover, one has
\be
E^{pp'}_k \ket{\Psi_0}
= \frac{r_{p'}}{r_p} E^{pp }_k \ket{\Psi_0} .
\ee
Therefore
\be
 V_\varphi h_k V_\varphi^{-1} \ket{\Psi_0} =
 \sum_{p,q=0}^{2 s} \sum_{p',q'=0}^{2 s}
\frac{r_{p'}r_{q'}}{r_p r_q} c^{pq}_{p'q'} c^{pq}_{p'q'}
 \rme^{i \varphi (q'-q)} E^{pp }_k E^{qq}_{k+1} \ket{\Psi_0}.
\ee
On the other hand,
\be
( F_{k+1} - F_{k}) \ket{\Psi_0} =
\sum_{p,q=0}^{2 s}
(f_q - f_p) E^{pp }_k E^{qq}_{k+1} \ket{\Psi_0}
\ee
Thus
\be
\label{CondLocalDivergenceComponentsgen}
 \sum_{p',q'=0}^{2 s}
\frac{r_{p'}r_{q'}}{r_p r_q} c^{pq}_{p'q'}
 \rme^{i \varphi (q'-q)} = f_q - f_p
\ee
determines the coupling constants of the spin-$s$ chain \eref{Hspin-s}.

This linear system of equations for the coupling constants
of the Hamiltonian can be easily solved which we
demonstrate for the first non-trivial case $s=1$. Notice that the case $s=1/2$
reproduces the $XXZ$-Hamiltonian discussed earlier.

\subsection{Spin-1 chain}

The ice rule \eref{icerule} allows for 19 non-vanishing coupling constants.
Hermiticity and reflection symmetry \eref{refsymmetry} leave
as free parameters the real-valued diagonal elements
$a_p := c^{pp}_{pp}$, $b_1 := c^{01}_{01} = c^{10}_{10}$,
$b_2 := c^{02}_{02} = c^{20}_{20}$, $b_3 := c^{21}_{21} = c^{12}_{12}$
and the spin-flip coefficients $c_1 := c^{01}_{10} = c^{10}_{01} \in \R$,
$c_2 := c^{02}_{20} = c^{20}_{02} \in \R$,
$c_3 := c^{12}_{21} = c^{21}_{12} \in \R$,
$d := c^{11}_{02} = c^{11}_{20}$, $\bar{d} := c^{02}_{11} = c^{20}_{11}$.
Requiring also spin-flip symmetry \eref{sfsymmetry}
leads to the further relations
$a_3=a_1$, $b_3=b_1$, $c_3=c_1$. Time-reversal symmetry then implies
$\bar{d}=d$.

\subsubsection{Computation of $h$ for helix states}

We define
\bel{defDeltazeta}
\delta = \cos{\varphi}, \quad \zeta = r_0 r_2/r_1^2.
\ee
The parameters $\varphi,\zeta$, or equivalently $\delta,\zeta$,
characterize the spin-1 helix state. In particular,
one has $\exval{S_k^x} = 2 \sqrt{2 \zeta} /(1+2\zeta) \cos{(\varphi (k-1))}$,
$\exval{S_k^y} = 2 \sqrt{2 \zeta} /(1+2\zeta) \sin{(\varphi (k-1))}$,
$\exval{S_k^z} = 0$, and the amplitude attains its maximum of full
polarization at $\zeta=1/2$.
We exclude from the discussion the non-helical
zero-current states $\varphi=0,\pi$ corresponding to $|\delta|=1$
and the non-helical states $\zeta = 0,\infty$ with vanishing
spin polarization $\exval{\vec{S}_k} = \vec{0}$.

The full set of equations (\ref{CondLocalDivergenceComponentsgen}) for the
spin-1 SHS reads
\begin{align}
a_0 & = a_2 =0\label{00gen}\\
b_1 &+c_1 \rme^{-i\varphi} + f_0-f_1=0\label{01gen}\\
b_1 &+c_1 \rme^{i\varphi} + f_1-f_0=0\label{10gen}\\
b_2 &+c_2 \rme^{-2i\varphi} + \bar{d} \zeta^{-1} \rme^{-i\varphi} + f_0-f_2=0\label{02gen}\\
b_2 &+c_2 \rme^{2i\varphi} + \bar{d} \zeta^{-1} \rme^{i\varphi} + f_2-f_0=0\label{20gen}\\
a_1 &+ d \zeta (\rme^{i\varphi} + \rme^{-i\varphi})  =0 \label{11gen}\\
b_3 &+c_3 \rme^{-i\varphi} + f_1-f_2=0\label{12gen}\\
b_3 &+c_3 \rme^{i\varphi} + f_2-f_1=0\label{21gen}.
\end{align}
Therefore
\bea
b_1 & = & - c_1 \delta \\
b_3 & = & - c_3 \delta
\eea
and $a_1 = - 2 d  \zeta \delta$,
$b_2 = - c_2 \cos{(2\varphi)} - \bar{d} \zeta^{-1} \delta$.

Since $b_2$ and $c_2$ are both real we conclude that also $d\zeta$
and $\bar{d} \zeta^{-1}$ must be real which implies that $d$ has the
negative phase of $\zeta$ plus a multiple of $\pi$.
For the coefficients $f_i$ one finds
\bea
f_0-f_1 & = & i c_1 \sin{\varphi} \\
f_1-f_2 & = & i c_3 \sin{\varphi}
\eea
In addition we have
\be
f_0-f_2 = i  c_2 \sin{(2\varphi)} + i \bar{d} \zeta^{-1} \sin{\varphi}
\ee
which yields the consistency condition $c_2 \sin{(2\varphi)} = (c_1+c_3-\bar{d} \zeta^{-1}) \sin{\varphi}$ which is automatically satisfied for
the irrelevant cases $\varphi=0,\pi$ and which
otherwise yields
\bea
d & = & \bar{\zeta} \left(c_1 + c_3 - 2 c_2 \delta \right) \\
b_2 & = & c_2 - (c_1+c_3)\delta \\
a_1 & = & 2 \delta |\zeta|^2 \left( 2 c_2 \delta - c_1 - c_3\right)
\eea
Thus all parameters are expressed in terms of $\zeta,\varphi$ characterizing the
helix state and the three
real-valued parameters $c_i$ that can be chosen freely.

With the shorthand $h_k \equiv h_k(c_1,c_2,c_3;\zeta,\varphi)$ we arrive at
\bea
h_k
\label{localH}
& = &
- c_1 \delta \left(E^{00}_k E^{11}_{k+1} + E^{11}_k E^{00}_{k+1}\right)
- c_3 \delta \left(E^{11}_k E^{22}_{k+1} + E^{22}_k E^{11}_{k+1}\right)
\nonumber \\
&  &
+ \left(c_2 -(c_1+c_3)\delta\right)
\left(E^{00}_k E^{22}_{k+1} + E^{22}_k E^{00}_{k+1}\right)
\nonumber \\
&  &
+ 2 \delta |\zeta|^2 \left( 2 c_2 \delta - c_1 - c_3\right) E^{11}_k E^{11}_{k+1}
\nonumber \\
&  &
+ c_1 \left(E^{01}_k E^{10}_{k+1} + E^{10}_k E^{01}_{k+1}\right)
+ c_3 \left(E^{12}_k E^{21}_{k+1} + E^{21}_k E^{12}_{k+1}\right) 
\nonumber \\
&  &
+ c_2 \left(E^{02}_k E^{20}_{k+1} + E^{20}_k E^{02}_{k+1}\right)
\nonumber \\
&  &
+ (c_1 + c_3 - 2 c_2 \delta) \left[ \zeta \left(
E^{01}_k E^{21}_{k+1} + E^{21}_k E^{01}_{k+1}\right)
+  \bar{\zeta} \left(
E^{10}_k E^{12}_{k+1} + E^{12}_k E^{10}_{k+1}\right)\right] .
\eea
We also note that
\be
F_{k} = f_1 \mathds{1} + (f_0-f_1) E^{00}_{k} - (f_1-f_2) E^{22}_{k}
= f_1 \mathds{1} + i \sin{\varphi} \left(c_1 E^{00}_{k} - c_3 E^{22}_{k}\right).
\ee
The constant $f_1$ is arbitrary since only the difference $F_{k+1} - F_{k}$
and the telescopic sum $\sum_{k=1}^{N-1} (F_{k+1} - F_{k}) =
F_{N} - F_{1}$ appear in calculations. Hence we can set $f_1=0$.

For spin-flip symmetry and time-reversal symmetry where $c_3=c_1$ and
$\bar{\zeta} = \zeta$ the local interaction reduces to
\bea
h^\ast_k(c_1,c_2;\zeta,\varphi)
\label{localH2}
& = &
- c_1 \delta \left(E^{00}_k E^{11}_{k+1} + E^{11}_k E^{00}_{k+1}+
E^{11}_k E^{22}_{k+1} + E^{22}_k E^{11}_{k+1}\right)
\nonumber \\
&  &
+ \left(c_2 -2c_1\delta\right)
\left(E^{00}_k E^{22}_{k+1} + E^{22}_k E^{00}_{k+1}\right)
\nonumber \\
&  &
+ 4 \delta \zeta^2 \left( c_2 \delta - c_1 \right) E^{11}_k E^{11}_{k+1} \\
&  &
+ c_1 \left(E^{01}_k E^{10}_{k+1} + E^{10}_k E^{01}_{k+1}
+E^{12}_k E^{21}_{k+1} + E^{21}_k E^{12}_{k+1}\right) \nonumber \\
& &
+ c_2 \left(E^{02}_k E^{20}_{k+1} + E^{20}_k E^{02}_{k+1}\right)
\nonumber \\
&  &
+ 2 (c_1 - c_2 \delta) \zeta \left(
E^{01}_k E^{21}_{k+1} + E^{21}_k E^{01}_{k+1}+
E^{10}_k E^{12}_{k+1} + E^{12}_k E^{10}_{k+1}\right) \nonumber
\eea
where $h^\ast_k(c_1,c_2;\zeta,\varphi) := h_k(c_1,c_2,c_1;\zeta,\varphi)$.
The corresponding divergence term is given by
\be
F_{k} =  i c_1 \sin{(\varphi)} \left(E^{00}_{k} - E^{22}_{k}\right)
= i c_1 \sin{(\varphi)} S^z_k.
\ee

\subsubsection{Integrable spin-1 chains with helix states}

The local Hamiltonian \eref{localH} is a special case of the family of spin-1 chains surveyed in \cite{Idzu94}.
For general parameter values the Hamiltonian built from the
local Hamiltonians \eref{localH} is not integrable which proves
that the phenomenon of ballistic transport in the helix state
is not related to integrability.
However, on a submanifold in parameter space one can identify two integrable families
which are special cases of the $U_q[\mathfrak{sl}(2)]$-symmetric Hamiltonian
\cite{Batc90}
\be
H^{BMNR} = \sum_{k=1}^{N-1} O_k(a,b;\lambda)
= \sum_{k=1}^{N-1} \tilde{O}_k(a,b;\lambda)
+ i a \sin(2\lambda) \left(S^z_{N} -  S^z_1\right)
\ee
where
\bea
\tilde{O}_k(a,b;\lambda) & = &
a \vec{S}_k \cdot \vec{S}_{k+1}
+ b \left(\vec{S}_k \cdot \vec{S}_{k+1}\right)^2 - (a+b)  \nonumber \\
& & i \frac{a+b}{2} \sin(\lambda)[(S^x_{k} S^x_{k+1}
+ S^y_{k} S^y_{k+1} +\cos{(\lambda)}S^z_kS^z_{k+1} )(S^z_{k+1} - S^z_{k}) + h.c.]  \nonumber \\
& & + 2(a-b) \sin^2(\lambda/2) [(S^x_{k} S^x_{k+1}
+ S^y_{k} S^y_{k+1}) S^z_{k} S^z_{k+1} + h.c.] \\
& & - \sin^2(\lambda) \left\{ 2a \left[ \left(S^z_k\right)^2
+ \left(S^z_{k+1}\right)^2 - 2\right] + \right.
\nonumber \\
&  & \left.
(a-b) \left[ S^z_kS^z_{k+1}
- \left(S^z_kS^z_{k+1}\right)^2 \right]\right\} \nonumber
\eea
with the spin-1 representation of $SU(2)$ and deformation parameter
$q=\rme^{i\lambda}$.

Comparing coefficients one finds
\be
h_k(c_1,-c_1,c_1,\frac{1}{\cos{(\varphi/2)}},\varphi)=c_1 \tilde{O}_k(1,-1,\varphi/2)
\ee
which is the integrable Zamolodchikov-Fateev Hamiltonian \cite{Zamo80}.
Moreover, one has
\be
h_k(0,c_2,0,\frac{1}{2\cos\varphi},\varphi) = \tilde{O}_k(0,c_2;1)
= c_2 \left[\left(\vec{S}_k\cdot\vec{S}_{k+1}\right)^2-1\right]
\ee
which is the bi-quadratic Hamiltonian of \cite{Klum89,Barb89}.
It is remarkable that there is no significant difference in the properties
of the helix states for the integrable and the non-integrable cases.
The integrable models, however, are of particular interest as they allow
for a more detailed study, including transport properties in the pure quantum
case and possibly the construction of non-local conserved quantities that are
relevant for the derivation of transport properties of these models
\cite{ProsIlievski13}.

\section{Concluding remarks}

We have  defined a family of spin helix states (SHS) with twist angle $\varphi$
in the $xy$-plane between neighboring spins and shown that these states arise
as the {\it exact} stationary solution of open spin-1 quantum chains with
bulk conservation of the $z$-component of the magnetization, but
boundary dissipation given by a suitably chosen two-parameter families of
Lindblad operators. These helix states are not in any sense close to the
quantum ground states of these spin chains. Nevertheless, they are stationary
under the Lindblad boundary driving that targets the boundary spins in different
directions, with a boundary twist angle $\Phi = (N-1)\varphi \mod{}2\pi $.
A non-zero winding number $K$ determined by $\varphi = (\Phi + 2\pi K)/(N-1)$
allows for a stationary spin-current $j^z$ of order 1.

Specifically, for the spin-1/2 Heisenberg chain with
anisotropy parameter $\Delta=\cos(\eta)$ the SHS occurs when $\eta=\varphi$.
As a function of $\eta$ the stationary current $j^z$ for fixed $\varphi$
shows a resonance-like peak at the SHS value $\eta=\varphi$. If this
matching condition is satisfied then
for any fixed anisotropy parameter $\Delta=\cos(\eta)$
the SHS carries a spin current $j^z = J \sin{(\eta)}$. This corresponds to
ballistic transport, i.e., the current does not depend on system size,
since for any $N$ one can find a boundary twist angle $\Phi \in [0,2\pi[$
that supports this current.
In fact, even when the boundary twist $\Phi$ is zero the SHS carries a current of
order 1 at anisotropies of the form $\Delta = \cos{2\pi K/(N-1)}$.
This is reminiscent of a result for the XXZ-chain with different Lindblad
operator where the
Drude weight has peaks at anisotropies $\Delta =
\cos{2\pi m/n}$ ($m,n$ being integers), leading
to an overall fractal behaviour of the Drude weight as a function of $\Delta$
in the thermodynamic limit $N\to\infty$ \cite{ProsIlievski13}.
Whether this Drude weight is related to an SHS is an open question.

We generalized the construction to higher spins.
For spin 1 we have derived Hamiltonians which allow for the existence
of stationary spin-1 SHS under suitable dissipative dynamics at the boundaries.
There Hamiltonians include the integrable Zamolodchikov-Fateev chain \cite{Zamo80}
and also the bi-quadratic Hamiltonian of \cite{Klum89,Barb89}.
We stress, however, that the existence of SHS is not in any way related to
integrability. Our solution includes non-integrable spin chains.
Moreover, since the construction relies on a local divergence condition
when applying the local Hamiltonian on the SHS, it can be generalized to any
lattice that allows for the cancellation of all these terms in the sum
of the local Hamiltonians over the lattice. So, in particular,
one can construct SHS for two- and three-dimensional cubic lattices.
By the same token, we expect that one can generalize the approach to
Hamiltonians with next-nearest neighour interactions and to Hamiltonians
with valence-bond eigenstates.

Generally, the properties of the SHS show, by comparing with known results
for other boundary driving mechanisms, that the transport properties of spin chains
depend qualitatively on the choice of Lindblad operators. This is somewhat
puzzling as the ballistic or other superdiffusive transport is expected
to be a bulk property of the chain, not a boundary property. This is
reminiscent of boundary-induced phase transitions in classical
stochastic particle systems \cite{Krug91,Popk99}.
Whether there is a deeper link is a further open question.

\section*{Acknowledgements}

Financial support by DFG is gratefully acknowledged. GMS thanks F.C. Alcaraz and
D. Karevski for stimulating discussions and the University of S\~ao Paulo and
the University of Lorraine for kind hospitality.


\begin{thebibliography}{99}
\bibitem{WindingNumber} We count the  winding number clockwise from $0$ to 
$N-2$ so that winding number $N-K-1$ corresponds to anti-clockwise winding number $K$.
\bibitem{spincurrent}
The spin current in \cite{Kare13b} differs from the spin
current considered here by a factor of two which arises from the
definition of the magnetization as $m=\exval{\sigma}/2$ as opposed to
$m= \exval{\sigma}$ in \cite{Kare13b}.
\end{thebibliography}
\end{document}